# Newtonian photorealistic ray tracing of grating cloaks and correlation-function-based cloaking-quality assessment

**Jad C. Halimeh,[1] Roman Schmied,[2] and Martin Wegener[3]**

[1]*Physics Department, Arnold Sommerfeld Center for Theoretical Physics, and Center for NanoScience, Ludwig-Maximilians-Universität München, D-80333 München, Germany*
[2] *Departement Physik, Universität Basel, Switzerland*
[3] *Institut für Angewandte Physik, DFG-Center for Functional Nanostructures (CFN), and Institut für Nanotechnologie, Karlsruhe Institute of Technology (KIT), D-76128 Karlsruhe, Germany*
[*]*jad.halimeh@physik.uni-muenchen.de*

**Abstract:** Grating cloaks are a variation of dielectric carpet (or ground-plane) cloaks. The latter were introduced by Li and Pendry. In contrast to the numerical work involved in the quasi-conformal carpet cloak, the refractive-index profile of a conformal grating cloak follows a closed and exact analytical form. We have previously mentioned that finite-size conformal grating cloaks may exhibit better cloaking than usual finite-size carpet cloaks. In this letter, we directly visualize their performance using photorealistic ray-tracing simulations. We employ a Newtonian approach that is advantageous compared to conventional ray tracing based on Snell's law. Furthermore, we quantify the achieved cloaking quality by computing the cross-correlations of rendered images. The cross-correlations for the grating cloak are much closer to 100% (*i.e.*, ideal) than those for the Gaussian carpet cloak.

©2010 Optical Society of America

## 1. Introduction

The concept of transformation optics [1-5] has firmly established that arbitrary objects can be made to appear invisible by surrounding them with an appropriately shaped inhomogeneous and generally anisotropic optical magneto-dielectric structure. To bring invisibility cloaking "down to earth", researchers now aim at obtaining good cloaking action for finite-size structures which are as small as possible compared to the object to be hidden, while obeying the following constraints: (i) avoid singularities of the optical properties, (ii) avoid metal structures necessary for magnetic resonances if possible (because of losses at optical frequencies), (iii) avoid very large anisotropies if possible (because this again requires resonances that lead to losses), and (iv) maintain operation over a wide frequency band.

The so-called carpet cloak [6] introduced by Li and Pendry has already come amazingly close to this aim. Corresponding experiments have been presented in Refs. 7-12. However, the carpet cloak does suffer from certain limitations, such as a pronounced lateral beam displacement [13]. This artifact – which was already visible in previous publications [6,14] – makes the cloaking structure detectable. We have recently pointed out [15] that the decay of the refractive-index profile away from the hidden object is related to the spatial-frequency distribution of the bump in the carpet. In particular, bump profiles with a cut-off at low spatial frequencies lead to a much faster decay of the refractive-index profile. Subsequently, the lateral beam displacement disappears almost completely for a single non-zero spatial frequency component. We have called this configuration the grating cloak [15].

However, it is difficult to assess the effect of the remaining imperfections of the finite-size cloaking structure from the mere inspection of *selected* rays in two dimensions (as done in our Ref. 15). In contrast, photorealistic ray tracing [16-18] is a well-known rendering technique based on ray optics, allowing to simulate images as they would appear to the human eye, and thus providing a direct and intuitive impression of their performance. These images can also be the basis for a quantitative assessment of the cloaking quality using cross-correlation functions (see section 5).

In this letter, we present corresponding numerical calculations. To obtain fully converged results under these conditions, we employ Newtonian ray tracing in a dedicated home-built software program.

## 2. The Grating Cloak

The refractive-index distribution $n$ of the grating cloak has been given [15] in closed form. Replacing the two-dimensional coordinates $(u,v)$ in equation (10) in Ref. 15 by $(x=u, y=v, z)$ in three dimensions, we have

$$n(x, y, z) = \frac{1}{\left|1 + W_0(icke^{ik(x+iy)})\right|}. \tag{1}$$

Here, $W_0$ is the Lambert function (or product logarithm). The constant wave number $k$ is given by $k=2\pi/\Lambda$ with the real-space grating period $\Lambda$; the complex-valued constant $c$ is given by $c=iA$ with real-valued $A$, where $2A$ is the peak-to-peak vertical grating amplitude (see (2)). Thus, numerical errors in determining the index profile (as, *e.g.*, possible for the quasi-conformal carpet cloak requiring numerical minimization of the modified-Liao functional) do not play any role at all in our present work. We do not consider any frequency dependence of $n$.

The profile $y(x)$ of the metal grating off of which light rays are reflected and which is to be cloaked by the refractive-index profile (1) has the shape of a trochoid given by the parametric form [15]

$$\begin{aligned} x(\xi) &= \xi - A\sin(k\xi) \\ y(\xi) &= A\cos(k\xi), \end{aligned} \tag{2}$$

where the parameter $\xi$ (corresponding to $x$ in Ref. 15) runs from $+\infty$ to $-\infty$. The grating profile (2) is translationally invariant along the $z$-direction and can be brought into the explicit form

$$x(y) = \pm\left(\frac{\Lambda}{2\pi}\arccos\left(\frac{y}{A}\right) - \sqrt{A^2 - y^2}\right) + N\Lambda, \tag{3}$$

with integer $N=-\infty, \ldots, +\infty$ for the different grating maxima. The "-" sign in (3) applies for the left-hand-side slope of each maximum, the "+" sign for the right-hand-side slope. In the shallow-grating limit defined by $2\pi A/\Lambda \ll 1$, hence $x \approx \xi$ in (2), one gets the approximate explicit form

$$y(x) \approx A\cos(kx). \tag{4}$$

However, for the numerical evaluation of the ray trajectories we will rather use the exact forms (2) or (3) throughout this work. Fig. 1(a) shows an example for the normalized grating-cloak model parameters $\Lambda=1$ and $A=3/(16\pi)\approx 0.06$ that lead to experimentally accessible refractive indices. In Figs. 3-7, one normalized unit corresponds to an actual 19 cm.

Since the grating cloak cannot be directly compared to the familiar carpet cloak [6], we also generate images for a single-bump Gaussian conformal cloak [15], which is more amenable to such a comparison. In the shallow-bump limit its bump shape is given by

$$y(x) \approx he^{-(x/w)^2}, \tag{5}$$

with the height *h* and the width *w*. However, for the actual calculations, we use the exact implicit form given in Ref. 15. Fig. 1(b) shows an example for the normalized model parameters *w*=0.306 and *h*=0.13.

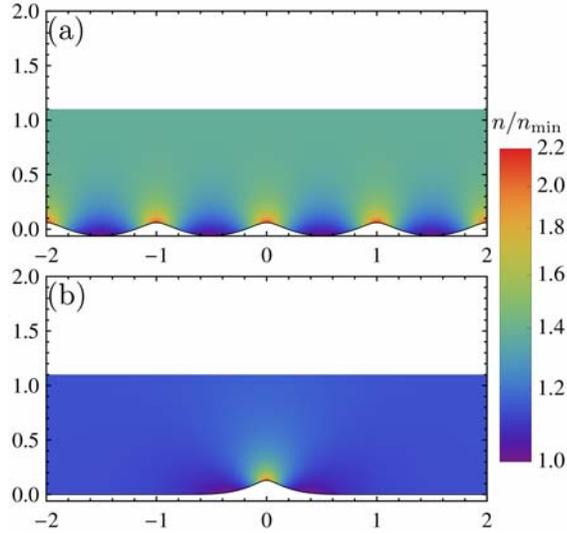

Fig. 1. (a) False-color representation of the normalized refractive-index profile $n/n_{\min}$ of the conformal grating cloak in the *xy*-plane according to (1). The normalized grating-cloak parameters are $\Lambda$=1 and $A$=3/(16$\pi$)≈0.06 (leading to $n_{\min}$=0.727). The normalized height of the cloak is equal to 1.1. In Figs. 3-7, one normalized unit corresponds to 19 cm. (b) Same, but for the Gaussian conformal map with *w*=0.306 and *h*=0.13 (leading to $n_{\min}$=0.873). The normalized height of the cloak is 1.1. Note that these parameters have been chosen such that $n/n_{\min}$ in (a) and (b) varies in the same interval [1.0, 2.2]. This aspect allows for comparing the two setups in a meaningful manner.

### 3. Newtonian Photorealistic Ray Tracing

We have previously [14,19] employed ray tracing based on Snell's law for sceneries involving dielectric cloaks. These results have been fully converged. However, later we noticed that certain viewing directions, such as a perpendicular view onto a cloaked bump, tend to converge only very slowly or not at all within accessible computation times. Sometimes, stripes appeared as artifacts in the rendered images. Thus, we started investigating alternative ray-tracing approaches.

It is well established [18,20-29] that ray optics in inhomogeneous optical media can be formulated in analogy to Hamiltonian mechanics. Here we use a simple and intuitive formulation that is equivalent to Newton's second law.

To emphasize the close analogy to mechanics and to aid the derivation, we briefly review three important equations (6)-(8) from classical mechanics. We start from Hamilton's variation principle for the action integral for the (generalized) coordinates and the (generalized) velocities for physical trajectories

$$\delta \int_{t_1}^{t_2} L(\vec{r}, \vec{v}, t) dt = 0. \tag{6}$$

As described in many mechanics textbooks, this allows for deriving the Euler-Lagrange equations for the components of the coordinate vector $r_i$ and the components of the velocity vector $v_i$

$$\frac{\partial L}{\partial r_i} - \frac{d}{dt}\frac{\partial L}{\partial v_i} = 0. \tag{7}$$

Identifying the first term as the force component $F_i$, the second term as the temporal derivative of the momentum component, and additionally assuming a constant particle mass $m$, one immediately arrives at Newton's second law for the acceleration

$$\frac{d\vec{v}}{dt} = \frac{d^2\vec{r}}{dt^2} = \frac{\vec{F}}{m}. \tag{8}$$

We can now simply repeat this procedure for light rays, where Hamilton's principle (6) has to be replaced by Fermat's principle, *i.e.*, by

$$\delta \int_{t_1}^{t_2} n(\vec{r}) |\vec{v}| \, dt = 0 \tag{9}$$

In analogy to (7), this leads to

$$|\vec{v}| \vec{\nabla} n - \frac{d}{dt}\left(n \frac{\vec{v}}{|\vec{v}|}\right) = 0, \tag{10}$$

and upon inserting the modulus of the local phase velocity of light $v = c_0/n$, which is known *a priori* at each point in space, to the Newtonian equation of motion

$$\frac{d\vec{v}}{dt} = \frac{d^2\vec{r}}{dt^2} = \frac{|\vec{v}|^2 \vec{\nabla} n - 2(\vec{\nabla} n \cdot \vec{v})\vec{v}}{n}. \tag{11}$$

The effective light-ray acceleration on the right-hand side of (11) vanishes in homogeneous media and thus outside the cloak. Inside the cloak, we solve this ordinary second-order differential equation (11) numerically for each ray. At the interface between the cloak and the surrounding air (or embedding medium), Snell's law and the Fresnel equations correctly describe the refraction and reflection of the ray due to the discontinuity in the refractive index. We neglect rays which are Fresnel reflected more than once, because their intensity will be negligible in practice.

The refractive-index profile of the grating cloak is given in the closed mathematical form (1). Evaluating the gradient yields

$$\vec{\nabla} n = \begin{pmatrix} \left(\frac{2\,\mathrm{Re}(1+W_0)\,\mathrm{Im}(W_0)}{|1+W_0|^2} - \mathrm{Im}(W_0)\right) \times \frac{k}{|1+W_0|^3} \\ \left(\frac{\mathrm{Im}^2(W_0) - \mathrm{Re}^2(1+W_0)}{|1+W_0|^2} + \mathrm{Re}(1+W_0)\right) \times \frac{k}{|1+W_0|^3} \\ 0 \end{pmatrix}, \tag{12}$$

with the abbreviation for the Lambert function $W_0$ (see (1))

$$W_0 = W_0(icke^{ik(x+iy)}). \tag{13}$$

Equation (12) can be inserted on the right-hand side of (11). We mention that, even though the refractive-index profiles used in this work are translation-invariant in the *z*-direction, and thus the *z*-component of the gradient of *n* is zero throughout space, the ray acceleration in the *z*-direction in (11) does in general not vanish.

## 4. Rendered Images

The scenery used for the following renderings is illustrated in Figs. 2(a) and 2(b). In essence, a model, a physicist, looks at herself in a curved mirror in front of her and observes the images depicted in Figs. 3-7. Technically, this means that a virtual point camera with a "human" *focal* field of view (*i.e.*, full opening angles of 50° horizontally and 42° vertically) is located between her eyes, which are centered in the photograph in Fig. 2(a). The distance between model and mirror is $d$=0.5 m (or 2.64 normalized units) – a typical distance for standing in front of your bathroom mirror. The model is 1.7 m (or 9.0 normalized units) tall. In the coordinate system shown in Figs. 1(a) and 1(b), the model is located at the top and looks downward onto the (a) grating or (b) Gaussian bump. For case (a), she sees three grating maxima within her vertical FOV: one at the center and two at the edges. For case (b), the single maximum is at the center of her FOV.

Corresponding photorealistic images rendered by the Newtonian ray-tracing approach described in the preceding section are shown in Fig. 3 for the grating profile and in Fig. 4 for the conformal Gaussian map. In both figures, (a) corresponds to a flat mirror and (b) to the corrugated mirror – a grating for Fig. 3 and a single bump for Fig. 4.

We see that panels (b) are distorted versions of panels (a). In both panels (b), a maximum of the profile is located in the vertical middle of the image. The maxima act as convex mirrors. This makes the model appear right-side-up in this part of the image, but strongly compressed along the vertical direction. For the grating mirror in Fig. 3, the minima of the profile act as concave mirrors. The next grating maxima appear at the very top and at the very bottom of the image in Fig. 3(b), respectively (*i.e.*, tan(42°/2)=Λ/$d$). Between maxima and minima, inflection points of the mirror profile occur. At each inflection point, the image turns from right-side-up to upside-down or *vice versa*. The combination of these aspects leads to highly complex and pronounced distortions with respect to the reference, Fig. 3(a).

In panels (c) of Figs. 3 and 4, a graded-index profile according to Fig. 1 is added onto the corrugated mirror profile. For an ideal cloak, panels (c) should be identical to panels (a), *i.e.*, all the distortions in the respective panels (b) should be compensated for. This is indeed very nearly the case on the vertical line through the center of the images, but increasing deviations occur towards the sides. The grating cloak in Fig. 3 appears to perform better than the Gaussian conformal map in Fig. 4. We will quantify these aspects by evaluating cross-correlation coefficients of panels (a) and (c) in the following section.

Figs. 3 and 4 remain unrealistic in that we have used the unreferenced refractive-index profile, *i.e.*, the index of air is taken as unity and the indices of the cloaks do exhibit values below unity, with the minimum values given in Fig. 1. This aspect makes experimental realization difficult. If one divides all refractive indices by the minimum value $n_{min}$, the results remain unchanged. However, it is more realistic to assume that we keep an air refractive index of $n$=1 (rather than $n$=1/$n_{min}$>1). It has been pointed out repeatedly [6-14] that this step introduces additional imperfections. These imperfections are visualized for the grating cloak in Fig. 5 (same cloak height as in Fig. 3, *i.e.*, 1.1 normalized units or 20.8 cm), in Fig. 6 (smaller cloak height of 0.55 normalized units or 10.4 cm), and in Fig. 7 (yet smaller cloak height of 0.41 normalized units or 7.8 cm). The "ideal" images as depicted in panels (a) of Figs. 5-7 are equivalent to adding a homogeneous dielectric plate with refractive index $n$=1/$n_{min}$>1 and with the same height as the respective cloak onto the flat mirror. These images are all different and *not* identical to Fig. 3(a) because the homogeneous dielectric plate with a refractive index larger than that of air effectively reduces the field of view. This effect can be seen by comparing the books at the upper right-hand side corner of Fig. 3(a) and Fig. 5(a). It decreases with decreasing dielectric-plate thickness from Fig. 5(a) via Fig. 6(a) to Fig. 7(a). Furthermore, one can see very faint Fresnel reflections in the high-resolution versions of panels (a) of Figs. 5-7 (see, *e.g.*, necklace on light blue shirt in Fig. 5(a)). The Fresnel transmissions also make the primary images very slightly dimmer than the reference image in Fig. 3(a).

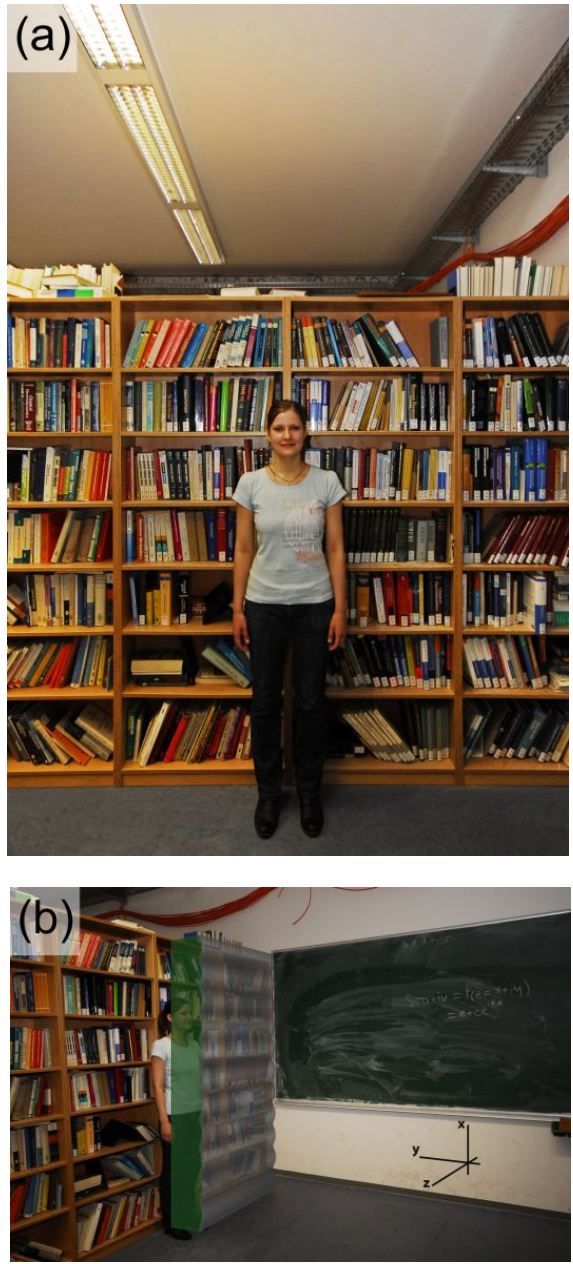

Fig. 2. Illustration of the scenery used for all photorealistic images rendered in this work. (a) This color image actually serves as the input for the following Newtonian ray-tracing calculations. (b) In these calculations, the model looks at the mirror in front of her (compare Fig. 1(a)) and observes the images shown in Figs. 3-7. The mirror (gray) is located at a distance of $d$=0.5 m from the model, the cloak is shown in green.

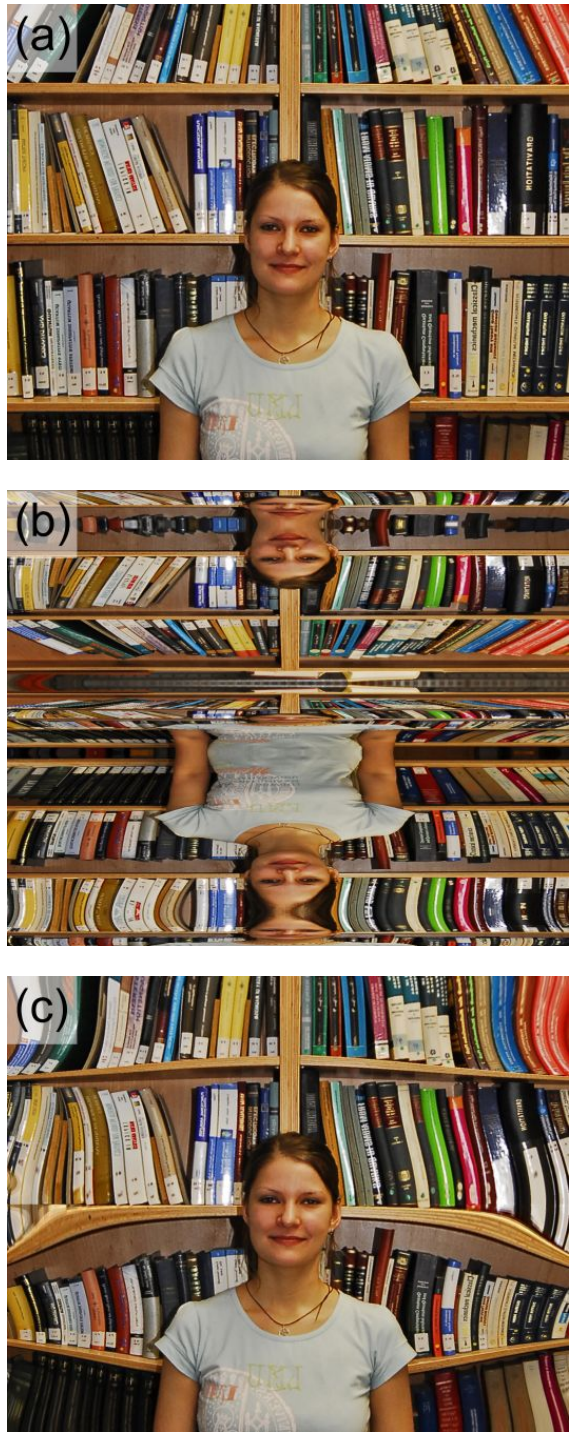

Fig. 3. Color images rendered by the Newtonian ray-tracing approach. The corresponding scenery is illustrated in Fig. 2. (a) Flat mirror. (b) Grating profile with $\Lambda=1$ (or 18.9 cm) and $A=3/(16\pi) \approx 0.06$ (or 1.1 cm), without cloak. (c) As (b), but *with* cloak. The cloak height is 1.1 normalized units (or 20.8 cm). Its refractive-index profile is depicted in Fig. 1(a). Note that this profile contains index values below unity with minimum $n_{\min}=0.728$.

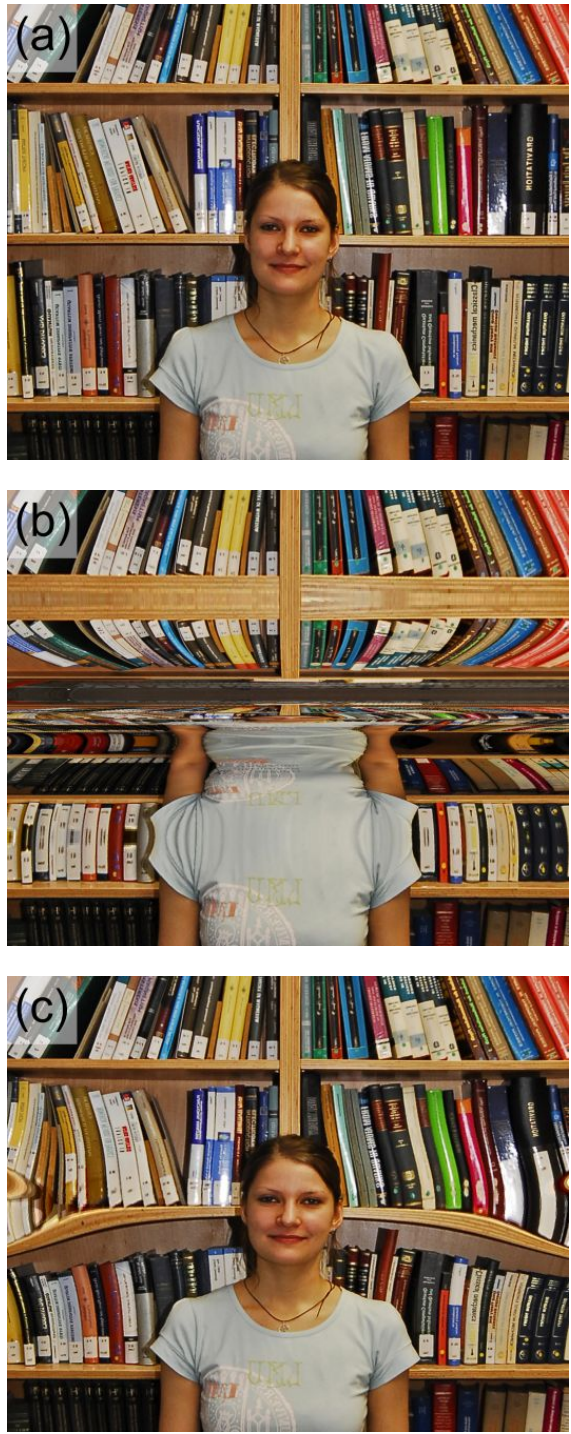

Fig. 4. As Fig. 3, but for the Gaussian conformal map with $w$=0.306 and $h$=0.13. The corresponding refractive-index profile is depicted in Fig. 1(b). Note that this profile contains index values below unity with minimum $n_{\min}$=0.873.

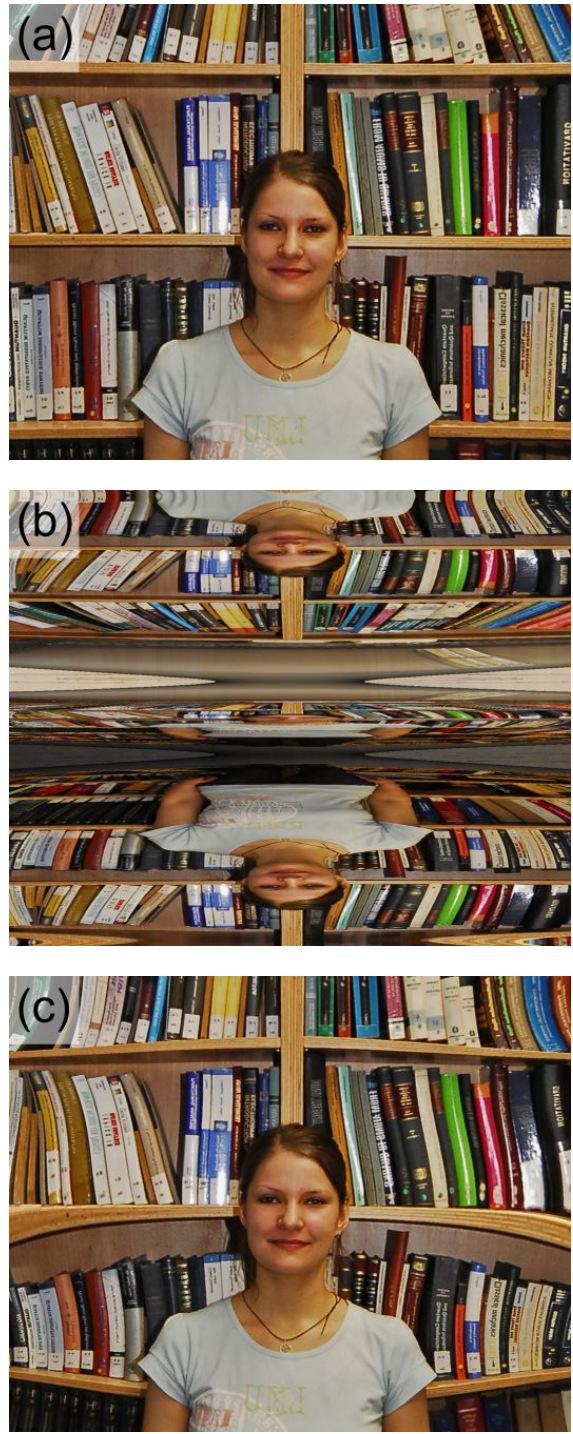

Fig. 5. As Fig. 3, but the refractive-index profile of the grating cloak is divided by $n_{min}$ (see Fig. 1(a)) such that the minimum refractive index in the cloaking profile becomes $n=1$. In (a) and (b), we have added a homogeneous dielectric plate with index $n=1/n_{min}=1.37$ and with the same height as that of the cloak (see Fig. 1(a)), *i.e.*, 1.1 normalized units (or 20.8 cm). The cloak height is successively reduced in Figs. 6 and 7.

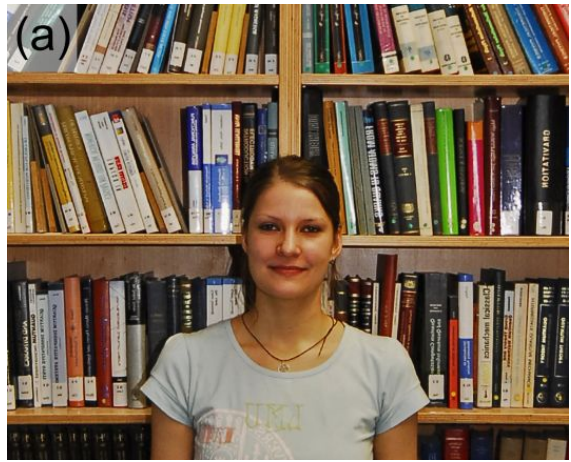

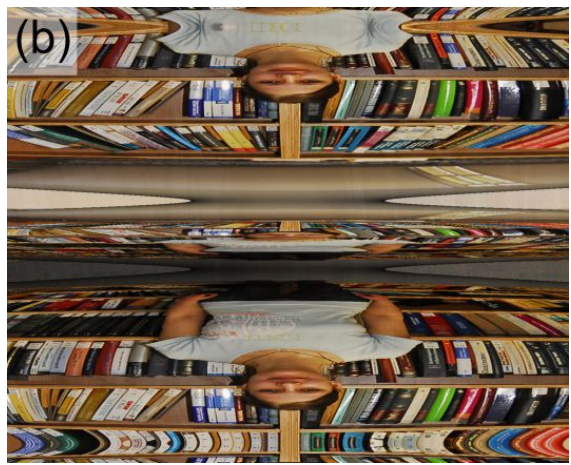

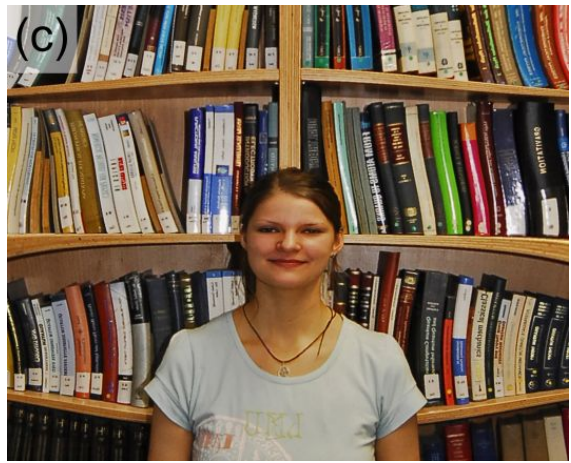

Fig. 6. As Fig. 5, but for a smaller height of the grating-cloak structure of 0.55 normalized units (or 10.4 cm).

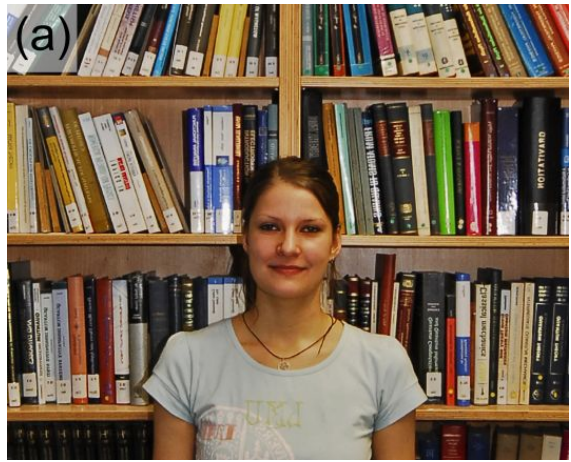
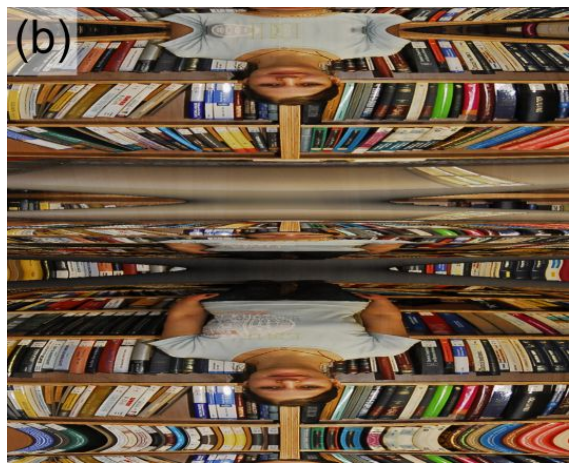
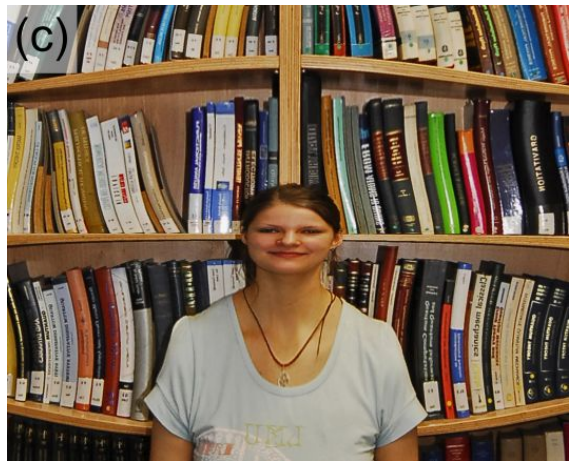

Fig. 7. As Fig 6, but for a yet smaller height of the grating-cloak structure of 0.41 normalized units (or 7.8 cm).

## 5. Correlation Function as Quantitative Measure of Cloaking Quality

Early work on electromagnetic cloaking [1-15] has been satisfied with some reasonable level of cloaking performance, judged by qualitative assessment. However, as the field of invisibility cloaking is getting more advanced, it is interesting to ask: Is there a quantitative measure for how good the cloaking performance of a particular cloaking device really is? With such a measure, one could quantitatively compare different cloaking approaches. Clearly there is no unique measure of this sort, but (cross-)correlation functions appear attractive to us as they have proven to be a useful approach in other fields of science and technology [30,31]. Let us briefly recall the underlying mathematics.

Suppose the functions $f(x,y) \geq 0$ and $g(x,y) \geq 0$ represent the intensity or gray levels of two monochrome two-dimensional finite-size photographs (like, *e.g.*, the images in Figs. 3-7) on an arbitrary scale, where $x$ and $y$ are the orthogonal coordinates in the photographic plane. (More generally, you can consider three such functions for the three principal colors in color photographs.) For example, let $f(x,y)$ be the unperturbed or reference image (*i.e.*, no object and no cloak) and $g(x,y)$ be the image taken with object and cloak. Thus, we will assume that the spatial average value of $g$ is not larger than the spatial average value of $f$. First, we subtract their respective spatial average values

$$\begin{aligned} f(x,y) &\to f(x,y) - \langle f \rangle \\ g(x,y) &\to g(x,y) - \langle g \rangle. \end{aligned} \quad (14)$$

Next, we define a two-dimensional image cross-correlation function, $C$, according to

$$C(\Delta x, \Delta y) := \frac{\iint f(x,y) g(x+\Delta x, y+\Delta y) dx dy}{\iint f^2(x,y) dx dy}. \quad (15)$$

For the case of ideal cloaking, *i.e.*, for $f(x,y)=g(x,y) \; \forall \; (x,y)$, we obtain the autocorrelation $C(0,0)=+100\%$. The value of $C$ will decay away from the maximum at $(\Delta x=0, \Delta y=0)$ towards 0% depending on the special properties of the photographic image $f(x,y)$. For example, periodic patterns in $f(x,y)$ will give rise to multiple maxima of $C(\Delta x, \Delta y)$. Non-periodic images $f(x,y)$ generally only lead to a single pronounced global maximum of $C(\Delta x, \Delta y)$ at $(\Delta x=0, \Delta y=0)$. If the image $g(x,y)=a \times f(x,y)$ with factor $0 \leq a \leq 1$ is just dimmer than the reference image $f(x,y)$, *e.g.*, due to some absorption or scattering of the cloaking device, we get $C(0,0)=a \leq 100\%$. If $g(x,y)$ is the "inverse" or complement (*i.e.*, black replaced by white and *vice versa*) of $f(x,y)$, we get $C(0,0)= -100\%$. If the image $g(x,y)$ is identical to the original $f(x,y)$, but shifted within the photographic $xy$-plane, a correlation maximum with a value approaching 100% (or strictly equal to +100% for infinitely large photographs) appears at some shifted position $(\Delta x \neq 0, \Delta y \neq 0)$. If $g(x,y)$ has nothing to do with $f(x,y)$, the correlation function will exhibit values near 0% for all $(\Delta x, \Delta y)$.

Thus, we could either declare $C(0,0)$ or the peak value of $C(\Delta x \neq 0, \Delta y \neq 0)$ as the requested single-number measure for cloaking quality. The former is stringent, the latter is more generous with respect to possible shifts in the images (like, *e.g.*, for the beam displacement of the carpet cloak). This shift could also be quoted. *Note that either of the two measures based on (15) takes advantage of the entire images and is not based on individual points of the images only.*

Ideally, one could finally average over the resulting cross-correlation functions for different reference images $f_i(x,y)$ and corresponding cloaked images $g_i(x,y)$ with index $i$ to eliminate any dependence on the particular image scenery.

Let us now apply the cross-correlation function (15) to the images shown in the previous section. The results are summarized in Table 1. There, "2D" refers to a near-zero field of view (FOV) in the horizontal direction, *i.e.*, to a vertical line through the center of the images in Figs. 3-7. Note that the grating cloak (like the carpet cloak) is exact in this two-dimensional (2D) world only. Indeed, under these conditions, the correlation is $C(0,0)$=-20% without cloak and as large as 99% with cloak – essentially perfect cloaking is achieved. Going to a three-dimensional setting reduces the value for the cloaked case to 91% for 10° FOV, to 73% for 25° FOV and to a mere 51% for 50° FOV; whereas, the values for the uncloaked case are all very small or even negative, indicating strongly distorted images.

Table 1. Cross-correlation coefficients $C:=C(0,0)$ according to (15) for different combinations of images, $f$ and $g$, as indicated in the two left-hand side columns for different horizontal field of views (FOV). FOV=0 corresponds to a vertical line in the middle of the images or to an effectively 2D situation, whereas increasing FOV corresponds to a more and more three-dimensional viewpoint. The full horizontal (vertical) width in Fig. 3 corresponds to a FOV=50° (FOV=42°) full opening angle. The rows in white are reference results for which the correlation should ideally be $C$=0%; the rows in light gray are cloaking results for which the correlation should ideally be $C$=100%. Cloaking action better than $C$=90% is highlighted in yellow.

| $f$ | $g$ | FOV=0° | FOV=10° | FOV=25° | FOV=50° |
|---|---|---|---|---|---|
| Fig. 3(a) | Fig. 3(b) | $C$=-20% | $C$=-1% | $C$=+7% | $C$=+6% |
| Fig. 3(a) | Fig. 3(c) | $C$=+99% | $C$=+91% | $C$=+73% | $C$=+51% |
| Fig. 4(a) | Fig. 4(b) | $C$=+17% | $C$=+46% | $C$=+42% | $C$=+43% |
| Fig. 4(a) | Fig. 4(c) | $C$=+90% | $C$=+87% | $C$=+73% | $C$=61% |
| Fig. 5(a) | Fig. 5(b) | $C$=+17% | $C$=+2% | $C$=+8% | $C$=+8% |
| Fig. 5(a) | Fig. 5(c) | $C$=+97% | $C$=+91% | $C$=+76% | $C$=+55% |
| Fig. 3(a) | Fig. 5(c) | $C$=+74% | $C$=+60% | $C$=+40% | $C$=+28% |
| Fig. 6(a) | Fig. 6(b) | $C$=+25% | $C$=+4% | $C$=+10% | $C$=+9% |
| Fig. 6(a) | Fig. 6(c) | $C$=+71% | $C$=+81% | $C$=+73% | $C$=+60% |
| Fig. 3(a) | Fig. 6(c) | $C$=+75% | $C$=+69% | $C$=+52% | $C$=+42% |
| Fig. 7(a) | Fig. 7(b) | $C$=+25% | $C$=+3% | $C$=+10% | $C$=+8% |
| Fig. 7(a) | Fig. 7(c) | $C$=+40% | $C$=+64% | $C$=+54% | $C$=+51% |
| Fig. 3(a) | Fig. 7(c) | $C$=+40% | $C$=+61% | $C$=+49% | $C$=+41% |

For the referenced cloaks in Figs. 5-7, one can either compare the cloaked images in Figs. 5-7 (c) with the flat mirror in air in Fig. 3(a) or with the flat mirror plus a dielectric plate in Figs. 5-7(a). The former comparison is stricter, the latter is more forgiving. The choice is ambiguous, however, because an observer will likely not be able to make a direct comparison between the two images. In any case, both options are listed in Table 1. Referencing to the flat mirror plus dielectric plate for the thick cloak in Fig. 5 leads to excellent correlations of 97% and 91% for 0° and 10° FOV, respectively. The correlation values are substantially lower when referencing to the flat mirror in air. The correlation values are yet smaller for the thinner cloaks in Figs. 6 and 7. Here, the lateral beam displacement [13,15], which depends on the angle of incidence of the rays onto the mirror and hence on the model's viewing direction, leads to large image distortions even for small vertical FOV and for referencing to the case with dielectric plate.

The Gaussian conformal map shown in Fig. 4 suffers from the same lateral-beam-displacement problem [15] – even for large cloak heights – leading to much lower correlation values in Table 1 than for the grating cloak of the same height (Fig. 3) despite the fact that only the middle part along the vertical of the image in Fig. 4(b) is actually distorted due to the single bump (compare Fig. 1). The latter fact can also be seen from the fairly large correlation values for the Gaussian uncloaked case (Fig. 4(b)). If one narrows the cross-correlation analysis down to the middle parts of Fig. 4(a) and (c) only, the corresponding $C$ further decreases.

## 5. Conclusion

We have visualized and quantified the performance of the grating cloak, which is an improved variation of the dielectric carpet cloak. While cloaking does exhibit cross-correlation coefficients near 100% (equivalent to perfect cloaking) for the finite-size grating cloak in 2D (or for small fields of view resembling 2D), the cross-correlation coefficients for the more realistic referenced 3D case with large fields of view are substantially smaller, although cloaking does appear reasonable at first sight. Our results should challenge others to obtain better cloaking as *quantified* by the cross-correlation coefficient under realistic conditions. We emphasize that the parameter choices in our work correspond to locally isotropic refractive indices between a minimum of 1.0 and a maximum of 2.2. These values are accessible throughout the entire visible spectral region with little dispersion, *e.g.*, by nanostructuring titania along the lines of Ref. 11.

## Acknowledgements

We thank our model, Tanja Rosentreter (LMU München), and the photographer, Vincent Sprenger (TU München), for help with the photograph used as input for the photorealistic ray-tracing calculations. We thank Manuel Decker (KIT Karlsruhe) for help in the processing of the figures. We acknowledge the support of the Arnold Sommerfeld Center (LMU München), which allowed us to use their computer facilities for our rather CPU-time-consuming numerical ray-tracing calculations. J.C.H. gratefully acknowledges financial support by the Excellence Cluster "Nanosystems Initiative Munich (NIM)". M.W. acknowledges support by the Deutsche Forschungsgemeinschaft (DFG), the State of Baden-Württemberg, and the Karlsruhe Institute of Technology (KIT) through the DFG Center for Functional Nanostructures (CFN) within subproject A1.5. The project PHOME acknowledges the financial support of the Future and Emerging Technologies (FET) programme within the Seventh Framework Programme for Research of the European Commission, under FET-Open grant number 213390. The project METAMAT is supported by the Bundesministerium für Bildung und Forschung (BMBF).